\documentclass[twocolumn,epjc3]{svjour3}

\usepackage{graphicx}
\usepackage{bm}
\usepackage{hyperref}
\usepackage{url}
\usepackage{microtype}
\usepackage{amsmath}
\usepackage{amssymb}
\usepackage{braket}
\usepackage{mathrsfs} 
\usepackage{bm} 
\usepackage{physics}
\usepackage{appendix}
\begin{document}

\title{Many-body approximations to the superfluid gap and critical temperature in pure neutron matter}

\author{Mehdi Drissi \thanksref{ad:triumf,ad:surrey,em:md}
\and
Arnau Rios \thanksref{ad:icc,ad:surrey,em:arh}}

\thankstext{em:md}{\email{mdrissi@triumf.ca}}
\thankstext{em:arh}{\email{arnau.rios@icc.ub.edu}}

\institute{
\label{ad:triumf}
TRIUMF, 4004 Wesbrook Mall, Vancouver, British Columbia V6T 2A3, Canada
\and
\label{ad:icc}
Departament de F\'isica Qu\`antica i Astrof\'isica,
 Institut de Ci\`encies del Cosmos (ICCUB), 
 Universitat de Barcelona,
Mart\'i Franqu\`es 1, Barcelona, E08028, Spain
\and
\label{ad:surrey}
Department of Physics, 
University of Surrey, Guildford GU2 7XH, United Kingdom
}

\date{Received: \today{} / Revised version: date}

\maketitle

\begin{abstract}
We compute singlet pairing gaps and critical temperatures in pure neutron matter with different many-body approximations. 
Medium effects tend to reduce gaps and critical temperatures compared
to the standard BCS ansatz. 
In the mean-field approximation, the ratio
of these two quantities remains constant  across a wide range of densities. 
This constant ratio is close to the universal prediction of BCS theory, 
whether three-neutron interactions are included or not. 
Using a more sophisticated
many-body approach that incorporates the effect of short-range correlations in pairing properties, 
we find that the gap to critical temperature ratio in the low-density regime
is substantially larger than the BCS prediction, independently of the interaction. 
In this region, our results are relatively close to experiments and theoretical calculations from the unitary Fermi gas. 
We also find evidence for a different density dependence of  zero-temperature gaps and critical temperatures
in neutron matter. 
\end{abstract}


\section{Introduction}\label{sec1}

Neutron stars (NSs) are key astrophysical objects. Their existence depends on several predicted properties of 
matter at the extremes of density, temperature and isospin asymmetry~\cite{Haensel,Shapiro}. In turn, the astrophysical
properties of NSs may be used to constrain the properties of extreme matter and, one hopes, the underlying nuclear physics properties.
The telltale example of how the connection between microscopic (nuclear) and macroscopic (astrophysical) properties 
can be established is the relationship between the Equation of State (EoS)
of nuclear matter and the mass-radius relation of NSs. 
General relativity is required to compute NS structure~\cite{Misner}, 
typically implemented by means of the hydrostatic, spherically symmetric and non-rotating Tolman--Oppenheimer--Volkov (TOV) equations. 
These equations require as input the EoS of dense matter, usually in the form of a nuclear physics inspired, one-to-one relation between energy
density and pressure~\cite{Steiner2013}. The solution of the TOV equations,
in turn, provides a mass-radius relation for NSs,
which can be immediately compared to observations~\cite{Lattimer2012,Ozel2016}. In the current era of multi-messenger astrophysics, there is an
ever increasing set of available NS data, including tens of pulsar masses from radio sources~\cite{Demorest2010,Cromartie2020,Fonseca2021};
a handful of radii from X-ray observations~\cite{Ozel2016,Lattimer2014,Raaijmakers2021}; and
tidal deformabilities from gravitational waves~\cite{Abbott2017,Abbott2018,Abbott2019}. 
Ultimately, the expectation is that a meaningful comparison between predictions and observations can constraint
the available space of EoSs and, with it, the underlying nuclear models
to generate them. 
As a starting step in this long-term endeavour, several EoS compilations are already available~\cite{Read2009,Oertel2017}, providing
a useful starting point for data-theory comparisons. The CompOSE database is one of such EoS compilations,
focusing (not exclusively) on finite-temperature effects for supernova simulations~\cite{Compose}.

While the mass-radius relation is a holy grail of NS physics, there are other aspects of NSs that are of general astrophysical interest.
Nuclear many-body theory indicates that neutrons undergo spontaneous pairing in the crust and outer core
of the star~\cite{Tamagaki1970,Pines1985}, much as they do in nuclei~\cite{Dean2003}. The existence of a 
superfluid phase may affect the dynamics of the star.
For example, the observation of period glitches of pulsars are often interpreted
as an abrupt transfer of angular momentum from superfluid matter to the rest of the star~\cite{Chamel2013}.
Similarly, the superfluid pair-formation
mechanism of relatively young pulsars may accelerate their cooling~\cite{Yakovlev2004,Page2011,Shternin2011,Elshamouty2013,Posselt2018,Wijngaarden2019,Wynn2021}. 
These observations provide an insight beyond the EoS of NS dense matter, with the tantalising potential to elucidate
different microphysical mechanisms at play in the interior of the star. 

Like the EoS, the superfluid properties of a NS are theoretical predictions based on a given set of ingredients and approximations. 
Several predictions of the pairing gaps are available in the literature~\cite{Dean2003,Gandolfi2015,Gandolfi2022}. 
Some of these have been reduced to simple numerical parametrizations for ease of practice~\cite{Ho2015}, although no universal
database is currently employed. 
In practice, 
in the astrophysical context, EoS and superfluid gaps are often exploited independently of their origin. In doing so, one may mix
together predictions that are based on very different nuclear physics
models. These inconsistencies are often disregarded, 
because the final astrophysical
observations may not always have enough resolving power to distinguish among different scenarios. 
But, as the quantity and quality of astrophysical
data influenced by superfluidity increases, a reduction of the parameter space is likely to provide more stringent constraints.
Theoretical consistency is imperative if the ultimate goal is to provide a propagation of information
from astrophysical observations all the way down to nuclear physics~\cite{Drischler2020}.

Here, we identify and quantify one specific obstacle in this direction.
More specifically, we examine an approximation employed in the astrophysical
context that is not necessarily justified in terms of nuclear many-body theory.
We look at the relation between the size
of the zero-temperature superfluid gap, $\Delta_0=\Delta(T=0)$,
and the critical temperature, $T_c$, of the superfluid transition.
Hereafter, we work with the choice of units $\hbar=k_B=1$.
In the simple Bardeen--Cooper--Schrieffer (BCS) model of superconductivity,
one finds that the ratio of these two quantities is
independent from the strength of the interaction, from the density of the medium
as well as from the cut-off of the regulator. In this case, the
universal ratio simply is $\Delta_0/T_c = \pi/e^\gamma \approx 1.764$~\cite{Abrikosov1965}. 
This remarkable result, or its corresponding extension to the triplet case~\cite{Hoffberg1970,Amundsen1985,Baldo1998,Khodel2004}, 
is exploited throughout the 
NS literature~\cite{Yakovlev2004,Ho2015}. 

In the following, we explore whether this BCS prediction of a constant $\Delta_0/T_c$ ratio remains valid for the singlet gap in the NS context. 
There are reasons to believe that this may not be the case. 
First of all, the impact of the medium on the pairing gap is sizeable and non-trivial in 
neutron matter~\cite{Dean2003,Rios2017a}. At the simplest possible level,
nucleon propagation is modified by a mean field potential. 
Beyond this mean-field picture, more sophisticated many-body approximations also predict
a substantial modification of single-particle strength.
Second, and equally relevant, the nucleonic interaction is far from the simple one
considered in the BCS model.
In particular, three-neutron forces (3NF) are important at nuclear densities, and 
their appearance is known to modify the pairing gap at zero temperature~\cite{Drischler2017}. 
We look at whether such modifications can affect the pairing gap, the critical temperature and their ratio. 

More fundamentally, the BCS many-body approximation can be interpreted as the lowest-order
of a many-body expansion in the context of superfluids.
Since the expansion parameter in this case is the strength of the potential,
one may question whether expectations from BCS predictions are reliable anymore in the strongly interacting, high-density regime of neutron stars. 
Along these lines, one can foresee two different types of medium modifications. In one instance, the gap $\Delta_0$ and the critical temperature 
$T_c$ may be modified, while keeping their ratio $\Delta_0/T_c$ intact. This probably requires some cancellations or fine tuning, which operate
at the BCS level and, possibly, beyond. Alternatively, the medium modifications may change both $\Delta_0$ and
$T_c$ in a way that their ratio is modified. This scenario has already been realised in several
instances in ultracold gases and high-$T_c$ superconductors, where the ratio $\Delta_0/T_c$ deviates substantially from the
BCS prediction~\cite{Magierski2011,Boettcher2014,Wu2019,Harrison2021,Drut2021}.

Chiral effective field theory interactions among nucleons provide a solid starting point to explore the properties of dense neutron matter in the region
of interest for singlet pairing superfluidity, up to and around saturation density~\cite{Epelbaum2009,Drischler2021}. Importantly, by looking at the
cutoff and scheme dependence, one can quantify to some extent the theoretical uncertainty of neutron-matter properties~\cite{Bogner2010}. 
Interactions are only
the starting point, though. For dense matter studies, one also needs to resort to a many-body scheme that provides an approximate solution for 
the thermodynamics and the superfluid properties. Many different methods exist to address specifically the EoS problem, both at zero and finite 
temperature~\cite{baldo1999nuclear,Muther2000,Baldo2012}. When it comes to superfluidity, however, one typically works in a
BCS approach, with some minor modification~\cite{Amundsen1985,Baldo1998,Khodel2001,Dean2003}. 
Extensions beyond the BCS level are typically phenomenological in 
nature~\cite{Shen2003,Shen2005,Cao2006}, 
or cannot easily be systematized to provide theoretical uncertainties~\cite{Schwenk2003,Schwenk2004,Ding2016,Rios2017a}.

Recently, we have formulated a new theoretical framework, dubbed Nambu-Covariant Green's Functions, that addresses this issue
and facilitates the systematic exploration of many-body effects in superfluid systems
at finite temperature~\cite{Drissi2021a,Drissi2021b}. This framework
develops a diagrammatic expansion that can potentially provide predictions for many-body scheme uncertainties. 
Importantly, the theory can also be used to build self-consistent approaches and hence provide a simplified superfluid counterpart to the more conventional self-consistent Green's function (SCGF) approach~\cite{Dickhoff2004,Rios2020}.
The numerical implementation of this Nambu-covariant Green's
function method is not yet available.
While this is developed, we explore here an alternative approach based on 
previous SCGF work~\cite{Muther2005,Ding2016,Rios2017a}.
We include the effect of short-range correlations (SRC)
effectively in the pairing gap by modifying the superfluid gap equation with convolutions of normal propagators. 
We study both the gap and the critical temperature calculations, 
to provide an initial assessment of the modifications of the $\Delta_0/T_c$ ratio in dense matter.
While this approach is not yet entirely consistent, we find a strong modification of the ratio
even in the low-density regime. Exploring the astrophysical implications of these findings will
require the full development of new techniques in the future.

We now proceed to discuss the two different many-body approximations that we employ
to compute the pairing gap, in an effort to provide
an estimate of theoretical uncertainties. 
We also provide parametrizations for the gap and the critical temperature dependence on Fermi momentum, 
in an attempt to assess whether these dependences are the same or not.
We first introduce the BCS approach with a Hartree-Fock (HF) single-particle spectrum, and discuss finite-temperature results.
Then, we discuss a finite-temperature extension of the SCGF model that can incorporate the effects of 
short-range correlations effectively in the superfluid phase. This extension was originally introduced in Ref.~\cite{Muther2005} and
further developed at zero temperature in Refs.~\cite{Ding2016,Rios2017a}. The results presented here represent
the first systematic exploration of this method across a wide range of temperatures.
 Our discussion is restricted to the singlet gap in the $^1$S$_0$ channel,
but the extension to an angle-averaged triplet case is relatively straightforward. 

\section{Bardeen-Cooper-Schrieffer pairing}

The momentum-dependent superfluid gap at a given temperature can be computed from an integral gap equation, which has the
generic structure
\begin{equation}
\Delta_k = -  \int \frac{\textrm{d}  k'}{(2 \pi)^3} 
\frac{\mathcal{V}(k,k')}{2 \chi_{k'} } \Delta_{k'} \, .
\label{eq:gap}
\end{equation}
Here, $\Delta_k$ is the singlet pairing gap that one is interested in, which appears in both sides of the equation. 
Within the BCS approximation, $\mathcal{V}(k,k')$ is the bare $NN$ interaction in the $^1$S$_0$ partial wave for the singlet pairing case of interest here.  
$k$ and $k'$ are, respectively, the incoming and outgoing relative momentum of a pair. The standard BCS kinematics 
involves a zero center-of-mass momentum~\cite{Dean2003}. 

The energy denominator $\chi_k$ contains information on the superfluid pairs, including the gap itself as well as
the temperature of the system. In a standard BCS approximation, the denominator is
\begin{align}
\frac{1}{2 \chi_k} = \frac{1-2 f(\xi_k)}{2 \xi_k} \, ,
\label{eq:chi_BCS}
\end{align}
with $f(\omega) = \left[ 1 + \exp \left( \frac{\omega-\mu}{T} \right) \right]^{-1}$ a Fermi-Dirac distribution.
The quasi-particle energies are defined by
\begin{align}
\xi_k = \sqrt{ (\varepsilon_k - \mu)^2 + \Delta_k^2 } \, ,
\label{eq:quasi}
\end{align}
where $\Delta_k$ denotes the gap contribution,
$\mu$ is the chemical potential,
and $\varepsilon_k$, the single-particle spectrum. 

Different approximations to $\varepsilon_k$ provide different pairing gaps. 
The standard BCS approach uses only a crude kinetic dispersion relation,
$\varepsilon_k=k^2/2m$. 
One step towards a more realistic description is to dress this dispersion relation with a mean-field $\Sigma_k$,
\begin{align}
\varepsilon_k = \frac{k^2}{2m} + \Sigma_k .
\label{eq:HF}
\end{align}
This is motivated theoretically by the well-known Hartree-Fock-Bogoliubov approach~\cite{RingSchuck1980}.
The mean-field incorporates the average effect of interactions in the medium,
 \begin{align}
\Sigma_k =  \int \frac{\textrm{d}  k'}{(2 \pi)^3} \langle \vec k \vec k' \lvert V \lvert \vec k \vec k' \rangle_A n_k' \, . 
\label{eq:spectrum}
\end{align}
Here, the interaction matrix elements 
are antisymmetrized and
$\vec k$ and $\vec k'$ represent single-particle momentum vectors. 
At the BCS level, the pairing interaction $\mathcal{V}(k,k')$ is related to the matrix elements
$\langle \vec k_1 \vec k_2 \lvert V \lvert \vec k_1' \vec k_2' \rangle_A$ 
by setting $\vec k_1=- \vec  k_2=\vec k$ and $\vec k_1'=- \vec k_2'=\vec k'$ and expanding to the 
appropriate partial wave. 

The momentum distribution in the superfluid case is given by the expression 
\begin{align}
n_k=1-\frac{\varepsilon_k - \mu}{\xi_k} \tanh \left( \frac{\xi_k}{2T} \right) .
\label{eq:momdis}
\end{align} 
At each Fermi momentum, $k_F=(3 \pi^2 \rho)^{1/3}$, we find the corresponding
chemical potential by inverting the expression
\begin{align}
\rho = 2  \int \frac{\textrm{d}  k}{(2 \pi)^3} n_k .
\label{eq:rho_mu}
\end{align} 
This already departs from the standard BCS approach, which employs $\mu=\varepsilon_F=k_F^2/2m$. 
This determination of $\mu$ is more 
consistent with the Hartree--Fock--Bogoliubov picture in the canonical ensemble, and with other
many-body approximations~\cite{Engelbrecht1997,Maly1999,Tsuchiya2009}. 
Compared to the standard BCS approach, this canonical implementation
yields a small reduction of the gap in the low-density regime. 

One of the key advantages of chiral interactions is that each order in the expansion gives rise to associated many-body
forces~\cite{Epelbaum2009}. These become more relevant as the density of the system increases. In the predictions shown here, we use the 
N3LO Entem-Machleidt (EM) interaction $V_{NN}$~\cite{Entem2003} and supplement it with a three-nucleon force (3NF) $W$ at N2LO. 
We use $\Lambda=500$ MeV in both cases.
The 3NF are incorporated following Refs.~\cite{Holt2010,Carbone2013a,Carbone2013b}, using 
the $P=0$ approximation and an internal nonlocal regulator.
We contract $W$ with the momentum distribution of Eq.~(\ref{eq:momdis}) to
obtain an effective two-body force, $\tilde W$. Carefully considering the
normal ordering of each contribution, one finds that in the calculation of the single-particle mean-field, 
$\Sigma$, one needs to employ the sum $V=V_{NN}+\frac{1}{2} \tilde W$. In contrast, in the pairing gap
$\mathcal{V}=V_{NN}+ \tilde W$~\cite{Drischler2017,Drissi2021a,Drissi2021b}. 
We compute the Hartree-Fock self-energy numerically with partial waves up to a total angular momentum 
of the neutron pair $J=10$. 

\begin{figure}[t!]%
\centering
\includegraphics[width=0.5\textwidth]{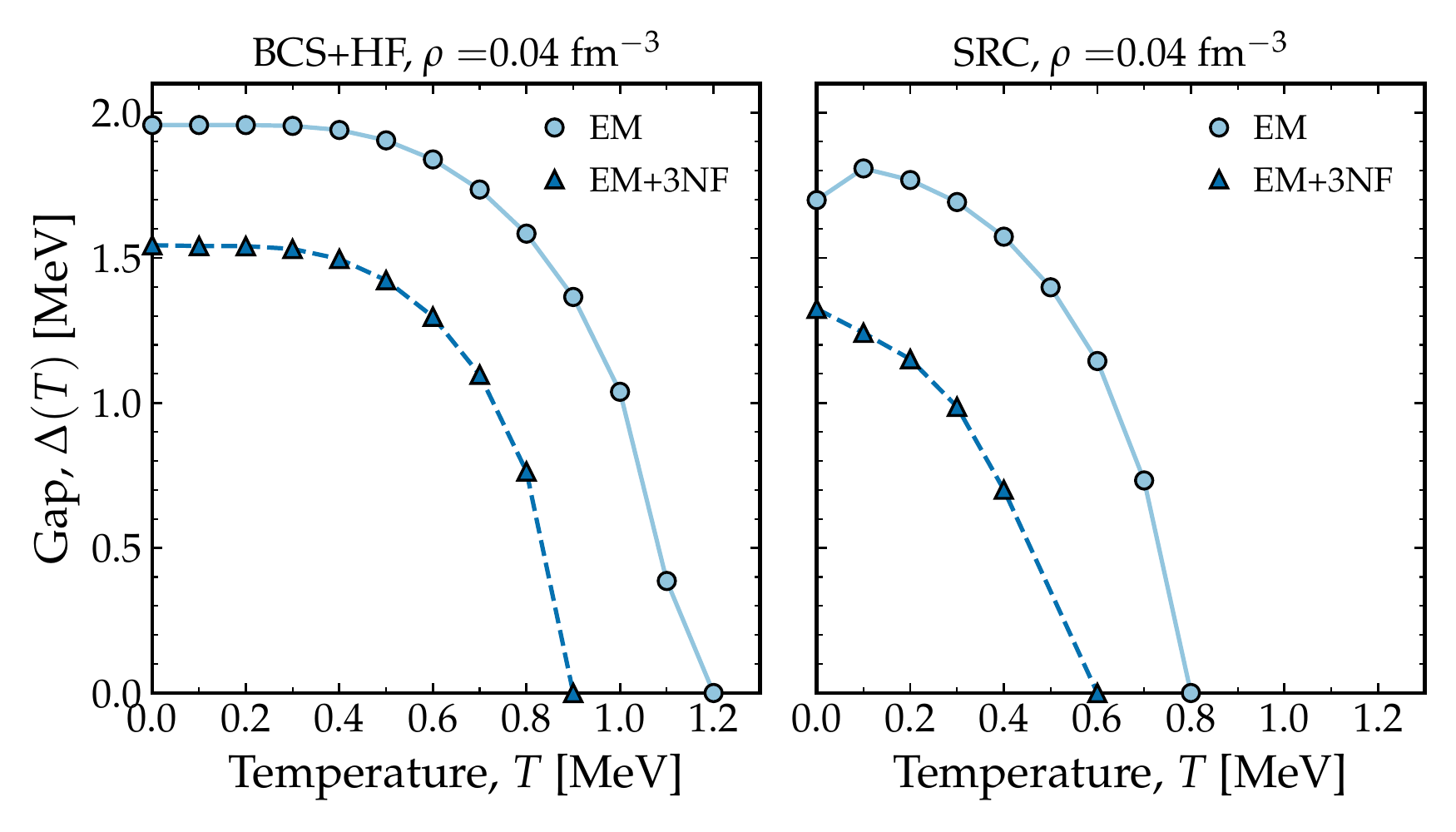}
\caption{
Left panel: superfluid gap $\Delta$ as a function of temperature $T$ in the BCS+HF approximation. 
Circles (triangles) correspond to results with the EM force (EM+3NF) force.
Right panel: the same for results using the SRC approximation. 
}\label{fig:gap_temperature}
\end{figure}

Because $\mu$ enters both the gap equation and the momentum distribution, the BCS+HF 
gaps and quasi-particle energies need to be computed iteratively. 
At each density and temperature, one can iterate Eqs.~(\ref{eq:gap})-(\ref{eq:HF})
to find a solution for the momentum dependent mean-fields, $\Sigma_k$, and gap functions, $\Delta_k$. 
As temperature increases, smearing out the Fermi surface, one expects the gap to decrease
until it disappears at a given $T_c$. We explore this phase transition by simulating several temperatures for a given
density (or Fermi momentum) value, as shown in Fig.~\ref{fig:gap_temperature}. 
The left panel shows the pairing gap at $k=k_F$, $\Delta(T)$,
for a density $\rho=0.04$ fm$^{-3}$ and varying temperatures, $T$. As temperature increases, the gap decreases monotonically
until it eventually disappears for a given critical temperature. The resolution of $T_c$ is limited by the numerical mesh. 
For the EM results only (full circles) shown in Fig.~\ref{fig:gap_temperature},  $T_c$ lies between $1.1$ and $1.2$ MeV. 
We thus estimate the critical temperature at this density as $T_c=1.15 \pm 0.05$ MeV. 
The inclusion of repulsive 3NF generally decreases the pairing gap. The EM+3NF results shown with triangles 
(dashed lines) are indeed lower across the whole temperature range. 
The pairing gap decays faster with 3NFs as a function of temperature, and the corresponding critical temperature is 
$T_c = 0.85 \pm 0.05$ MeV. 
We stress that these
results incorporate a temperature-dependent single-particle spectrum $\Sigma$, with 3NFs 
that are consistently included in the pairing interaction. 
This complicates the potential application of alternative methods that estimate $T_c$
using algorithms relying on the Thouless criterion,
like the Weinberg eigenvalue approach used in Ref.~\cite{Ramanan2013}.

\begin{figure}[t!]%
\centering
\includegraphics[width=0.4\textwidth]{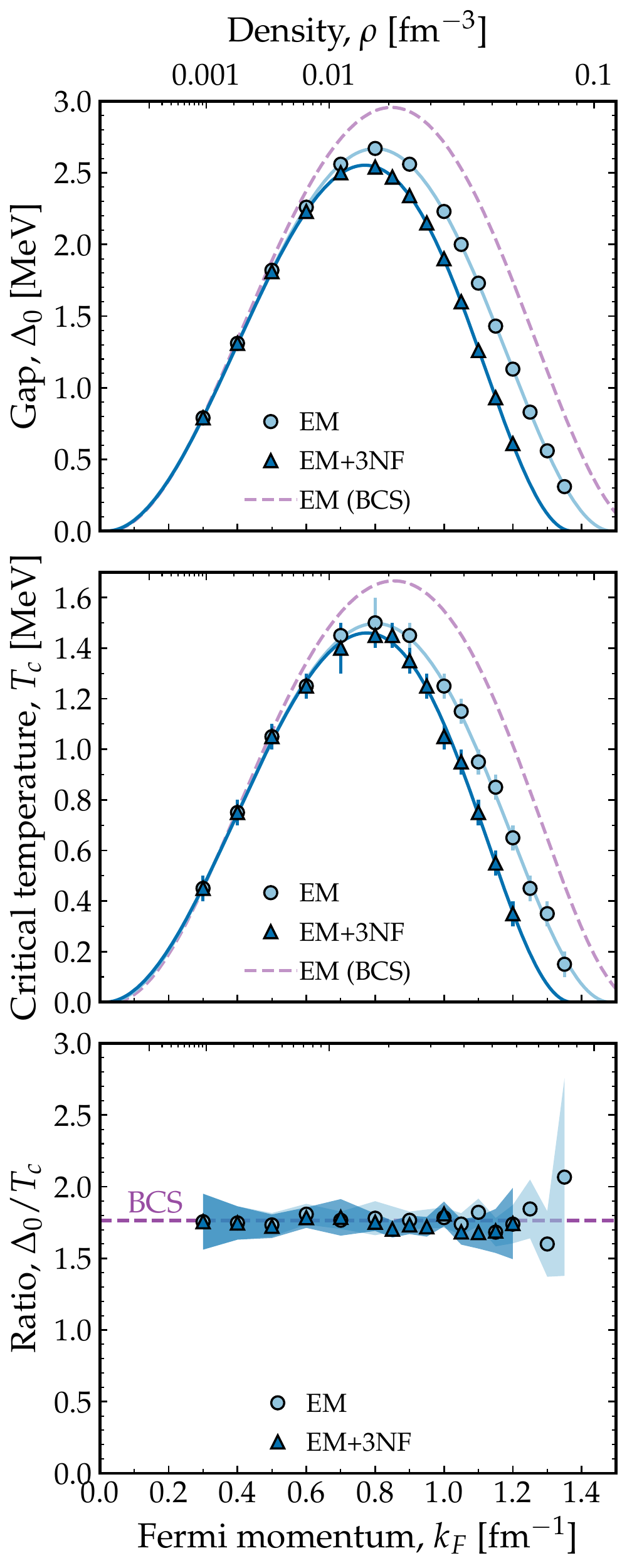}
\caption{
Top panel: zero-temperature superfluid gap $\Delta_0$ as a function of neutron matter Fermi momentum. 
Filled circles (triangles) represent the BCS+HF solution for the N3LO EM (N3LO+3NF) potential. 
Dashed lines correspond to the BCS solution for the EM interaction. 
Central panel: critical temperature $T_c$ as a function of neutron matter Fermi momentum. 
The error bars represent the numerical uncertainty in the determination of $T_c$. The symbols have the same meaning as the the top panel. 
Bottom panel: ratio of $T_c$ to $\Delta$ as a function of Fermi momentum. The dashed horizontal line is the BCS result $\Delta_0/T_c=1.764$.
}\label{fig:tc_gap_ratio_bcs}
\end{figure}

One can proceed similarly for different densities, generating a mesh of gap values for different thermodynamical conditions. 
With this, one gets the results displayed in Fig.~\ref{fig:tc_gap_ratio_bcs}. The top panel shows the zero-temperature gap, $\Delta_0$,
as a function of Fermi momentum.
We show three different sets of results. The dashed line is a reference 
standard BCS value, obtained with the EM NN interaction only; a non-interacting
single-particle spectrum, i.e. $\Sigma=0$; and a chemical
potential $\mu=\varepsilon_F$. In contrast, the circles and triangles joined by a full line correspond to EM BCS+HF 
results ($\Sigma \neq 0$) with (triangles) or without (circles) 3NFs.
In these results, the chemical potential is determined using Eq.~(\ref{eq:rho_mu}).

The gaps show the standard dumbbell shape for all cases. The position and size of the maximum gap depends on the many-body scheme. 
For the EM BCS results, the gap maximum is close to $3$ MeV at $k_F=0.9$ fm$^{-1}$. 
The BCS+HF results have in general lower gap values, with a maximum around $\approx 2.7$ MeV at a Fermi
momentum of $\approx 0.8$ fm$^{-1}$. As density increases beyond the peak, the gap subsequently decreases and closes near $1.4-1.5$ fm$^{-1}$
for the BCS+HF simulations, and around 1.6 fm$^{-1}$ for the BCS-only case. 
The effect of the 3NF is to reduce the maximum gap by about $0.2$ MeV, and also to bring the gap closure to lower densities.
This effect is well-known and has already been identified in BCS and in BCS+HF calculations~\cite{Rios2017a,Drischler2017}. 
Our results also coincide with those generated by Drischler \emph{et al.} in Ref.~\cite{Drischler2017}. 

While the symbols in Fig.~\ref{fig:tc_gap_ratio_bcs} represent the numerical simulations, the lines have been
obtained by fits to the data with the function 
\begin{align}
g^x(k_F) = g^x_0
\frac{ (k_F-k^x_0)^2}{(k_F-k^x_0)^2+k^x_1} 
\frac{ (k_F-k^x_2)^2}{(k_F-k^x_2)^2+k^x_3}
 \, ,
\label{eq:gapfit}
\end{align}
with $g^x_0$, $k^x_0$, $k^x_1$, $k^x_2$ and $k^x_3$ numerical parameters \cite{Ho2015}. Here, $x$ labels the data that is fitted, with $x=\Delta_0$ for pairing gaps and $x=T_c$ for critical temperatures. 
Details on the numerical fit to this function are given in Appendix~\ref{app:fits}. 
We note that the fits provide a very good reproduction of the BCS results, 
with maximum deviations with respect to data of $0.04$ MeV across the whole temperature regime. 
The top rows of Table~\ref{table:BCS} provide the corresponding fit parameters for the pairing gaps.
$k^\Delta_0$ and $k^\Delta_2$ represent the Fermi momenta at which the gap
 opens and closes, respectively. The opening Fermi momentum is the same for both the EM and 
 EM+3NF calculations, as expected naively from the growing importance of 3NFs with density.
 In contrast, the fit indicates that the closing Fermi momentum for the singlet gap without 3NFs,
 $k_2^\Delta=1.49$ fm$^{-1}$, is larger than the EM+3NF result, $k_2^\Delta=1.37$ fm$^{-1}$.

\begin{table}[t!]
\caption{Parameters generated by a fit to the calculated gaps, $\Delta_0$ (top 2 rows), and the critical 
temperatures (bottom 2 rows). These results correspond to the BCS+HF approximations. \label{table:BCS}}
\centering 
\begin{tabular}{cccccc} 
\hline \hline 
BCS+HF & $g_0^\Delta$ & $k_0^\Delta$ & $k_1^\Delta$ & $k_2^\Delta$ & $k_3^\Delta$ \\
Units & MeV & fm$^{-1}$ & fm$^{-2}$ & fm$^{-1}$ & fm$^{-2}$ \\
 \hline
EM & 24.76 & 0.02 & 1.23 & 1.49 & 0.89 \\ 
EM+3NF & 15.09 & 0.02 & 1.02 & 1.37 & 0.62 \\ \hline 
 & $g_0^{T_c}$ & $k_0^{T_c}$ & $k_1^{T_c}$ & $k_2^{T_c}$ & $k_3^{T_c}$ \\
 \hline
EM & 7.30 & 0.02 & 0.92 & 1.48 & 0.69 \\ 
EM+3NF & 7.94 & 0.01 & 1.00 & 1.37 & 0.59 \\ \hline 
\hline 
\end{tabular} 
\end{table}
 
The central panel in Fig.~\ref{fig:tc_gap_ratio_bcs} shows $T_c$ as a function of Fermi momentum, 
obtained as explained
earlier by computing the gap at different temperatures. 
Because of the limited numerical mesh of temperatures, we assign an error to $T_c$, 
which is typically $\epsilon_{T_c} =0.05$ MeV or, on occasions, $0.1$ MeV. The
data has a very similar shape to the BCS+HF zero-temperature gap - a dumbbell, with a maximum of $\approx 1.5$ MeV
at around $k_F \approx 0.8$ fm$^{-1}$. $T_c$ decreases at high
density, until it closes around $1.4$ fm$^{-1}$. 
BCS results without HF spectra (dashed lines) predict substantially larger critical temperatures. 
To quantify the 
similarity between the shape of the gap 
parameter and the critical temperature, 
we fit the function of Eq.~(\ref{eq:gapfit}) to the $T_c$ data, considering
the theoretical uncertainty. 
We provide the fit parameters in the bottom rows of Table~\ref{table:BCS}. Looking again at the
opening and closing Fermi momenta,  $k_0^{T_c}$ and $k_2^{T_c}$, we find a very good agreement
between the gap and the critical temperature data. 
The comparison between the values of 
$k_1^x$ and $k_3^x$ can only be done in relative terms. 
We find that the ratio $k_1^x/k_3^x$ for the gap and $T_c$ fits with the EM interaction are close to each other,
$k_1^x/k_3^x=1.3-1.4$. 
For the EM+3NF fits, the fits provide two consistent ratios 
$k_1^x/k_3^x =1.6-1.7$, which are different to the EM ones.
These ratios illustrate the faster decline with Fermi momentum of the EM+3NF 
results as opposed to the EM-only ones.
With this in mind, we conclude that the gap and the critical
temperature are different for different interactions, but
they have essentially the same dependence in $k_F$ within the BCS+HF approximation.

Finally, the bottom panel of Fig.~\ref{fig:tc_gap_ratio_bcs} shows the results for the ratio of the
gap to the critical temperature values, 
$\Delta/T_c$. The bands correspond to the (independent) 
uncertainty propagation of 
the errors associated to the gap (negligible at the BCS+HF level) and the critical temperature
(of order $\epsilon_{T_c} =0.05$ MeV, as stated above). We raise at least two relevant
astrophysical conclusions from this result. First, the ratio is obviously quite constant over a
wide density regime. This supports the statement in the previous paragraph
that the functional forms of $\Delta_0(k_F)$ and
$T_c(k_F)$ are very similar in the BCS+HF approximation.
Second, the ratio does coincide with the BCS result, 
$\Delta_0/T_c=1.764$~\cite{Abrikosov1965}
within the uncertainties generated by our results which are of the order
of $0.1$. The deviations from
the universal constant are commensurate with this uncertainty 
and there are no clear
trends nor departures from this value.

We stress that the results presented here, within the BCS+HF approach (solid lines), go beyond the standard
BCS (dashed line) approximation that is used to derive the identity $\Delta_0/T_c=1.764$. 
This result arises from an analysis of the kernel of Eq.~(\ref{eq:gap}) close to $T=0$ and close to $T_c$. 
The standard analysis 
neglects the effects of the momentum dependence (or the non-locality) of the interaction, 
and is performed in the absence of 3NFs. Likewise, the analysis typically neglects the 
 single-particle mean-field, $\Sigma=0$.
Here, instead, $\Sigma$ is momentum, density and temperature
dependent at the HF level. 3NFs affect the pairing properties both through $\Sigma$ itself,
but also through the pairing interaction, $\mathcal{V}$. And, yet, even though these
changes do substantially modify the values of
 $\Delta_0$ and $T_c$ independently (see top and central panels), 
 our results indicate that the ratio
$\Delta_0/T_c$ is independent of all such medium modifications.

\section{Beyond-BCS pairing}

We now apply the very same methodology in an approximation that goes beyond the BCS+HF approximation. 
We aim to test whether the ratio $\Delta_0/T_c$ can be modified, but we also want to quantify the deviation from the BCS+HF results. 
To this end, we implement an approximation that incorporates the fragmentation of quasi-particle strength
effectively in the pairing properties~\cite{Muther2005,Ding2016,Rios2017a}. 
The single-particle spectral function, $A_k(\omega)$, provides a positive-definite probability
distribution which describes the fragmentation of strength for a given momentum, $k$,
in the normal state~\cite{Dickhoff_book,vanLeeuwen_book}. 
In a mean-field approximation, such as HF, this is just a $\delta$ peak centered
at a certain effective single-particle energy $\varepsilon_k$.
Going beyond mean-field approximations, however, the
spectral functions departs from this basic $\delta$
distribution~\cite{Dickhoff_book,vanLeeuwen_book}. 

In the SCGF approach
at finite temperature, $A_k(\omega)$ is instead a density and temperature-dependent function, 
with a strong peak around a preferred single-particle energy, but also relatively large widths
beyond the Fermi surface~\cite{Rios2020}. 
We access the spectral function by means of finite temperature SCGF calculations~\cite{Carbone2013a,Rios2017a,Rios2020}.
For any given density and temperature, a SCGF simulation sums an in-medium (particle-particle and hole-hole) ladder series at finite temperature~\cite{Bozek1999,Frick2003}.
 For a fixed $\rho$, we can perform simulations for several values of $T$, 
down to minimum temperatures of order $2$ MeV, well
above any pairing instability~\cite{Ramos1989}. 
With this data, we can extrapolate below the minimum temperature value and obtain a normal spectral function 
in a regime which may be superfluid, as explained in Ref.~\cite{Ding2016}. 
This extrapolation allows us to compute a fragmented pairing contribution
to the denominator,
\begin{equation}
\frac{1}{2 \zeta_k} = \int  \frac{\textrm{d}  \omega}{2 \pi}  \frac{\textrm{d}  \omega'}{2 \pi}  \frac{1-f(\omega)-f(\omega')}{\omega+\omega'}  A_k (\omega) A_k (\omega') \, ,
\label{eq:fragden_normal}
\end{equation}
so that $\chi_k=\sqrt{\zeta^2_k+\Delta^2_k}$ in the pairing gap equation, Eq.~(\ref{eq:gap})~\cite{Muther2005}. 
We note that the extrapolation procedure only works for systems which are relatively degenerate or, in other words, for relatively large
densities, above $k_F \approx 0.6$ fm$^{-1}$. 

The extrapolated pairing denominator $\chi_k$ can be directly employed in the gap equation, Eq.~(\ref{eq:gap}). 
We stress, however, that many-body 
theoretical approaches indicate that the integral convolution should also include one superfluid (as opposed to normal) spectral function~\cite{Bozek2000}.
 In other words, we are missing feedback effects from the superfluid into the gap itself. These effects occur very close
 to the Fermi surface and are expected to be relatively small~\cite{Muther2005}. 
In the absence of the full implementation of a consistent many-body theory of superfluids, as presented in Refs.~\cite{Drissi2021a,Drissi2021b},
we use this approach as an initial step forward in the analysis. 

\begin{figure}[t]%
\centering
\includegraphics[width=0.4\textwidth]{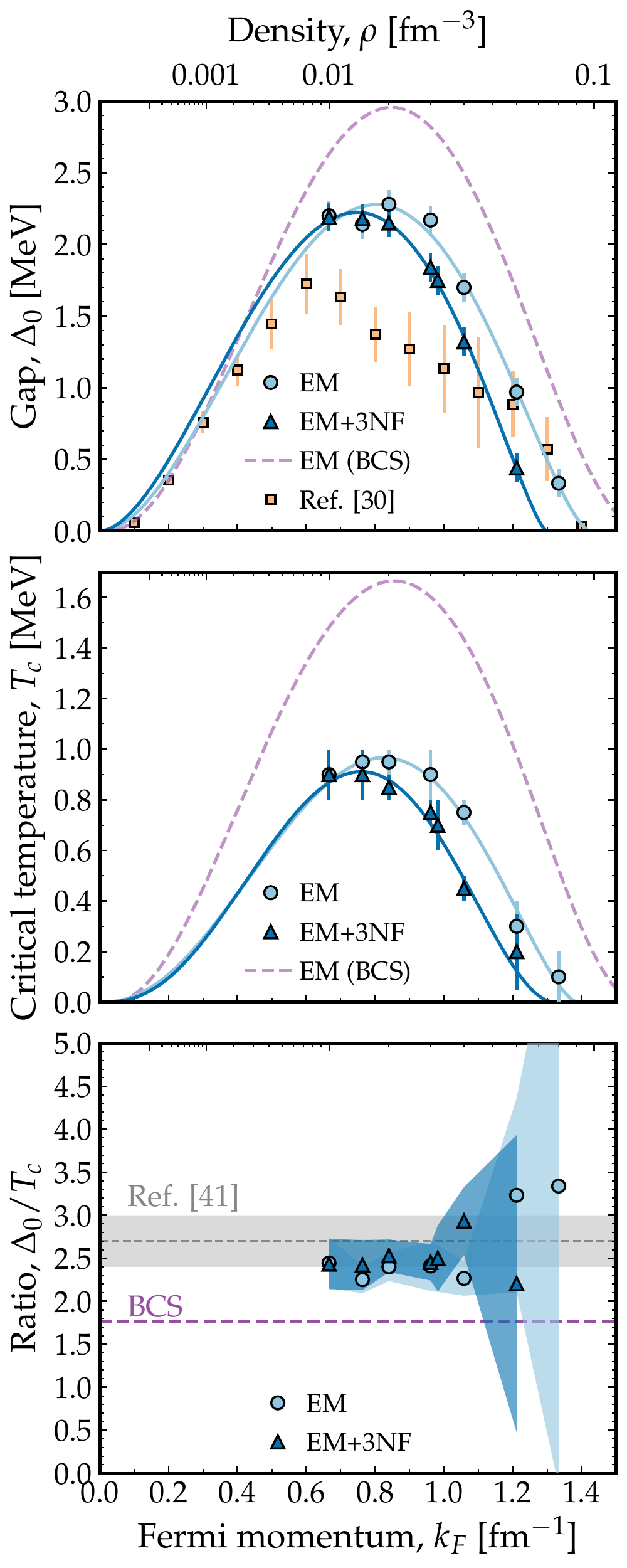}
\caption{
The same as Fig.~\ref{fig:tc_gap_ratio_bcs} for the SRC approximation.
}\label{fig:tc_gap_ratio_src}
\end{figure}

SRC associated with realistic NN forces induce fragmentation on spectral functions. This fragmentation
reduces the pairing gap~\cite{Bozek1999,Muther2005}. 
Our simulations corroborate this conclusion at finite temperatures. The right panel of Fig.~\ref{fig:gap_temperature} shows SRC pairing
gaps as a function of temperature for EM (circles) and EM+3NF (triangles). 
The SRC gaps are smaller than their BCS+HF counterparts
across the whole density regime. Importantly, the temperature dependence of these results is steeper, and the corresponding critical temperatures
are substantially reduced. We note that the $T=0$ temperature result for the EM interaction deviates from the monotonous 
 dependence obtained with the finite-temperature results. This is most likely an artefact of the numerical extrapolation procedure. In the following,
 we associate a constant theoretical uncertainty, $\epsilon_{\Delta_0}=0.1$ MeV, to our results, to account for such issues. 
In the case of EM+3NF, this error also incorporates the uncertainty associated to the fact that the
reduction of the 3NF into an effective 2-body force is performed with a step-function momentum distribution, $n_k$, as opposed to the pairing-corrected distribution of Eq.~(\ref{eq:momdis}).

 We present the results for $\Delta_0$ in the SRC for the EM (circles) and the EM+3NF
(triangles) interactions in the top panel of Fig.~\ref{fig:tc_gap_ratio_src}. This figure also includes the BCS results (dashed line).
Compared to Fig.~\ref{fig:tc_gap_ratio_bcs}, the SRC results provide smaller gaps across the whole density regime, 
with maximum gaps here
of order $2.4$ MeV. Just as in the BCS+HF case, 3NFs tend to reduce 
the maximum gap
and the closure Fermi momentum  further. We provide
the associated fits to this data in Table~\ref{table:SRC}. 
These fits suggest that the gap opens at zero Fermi momentum, $k_0^\Delta=0$ fm$^{-1}$. This is the lowest bound 
imposed in our fitting procedure.
We stress,
however, that the lack of SRC data in the region below the peak does not help in constraining the low-density parameters. 
The closure gap Fermi momentum, $k_2^\Delta$, 
in contrast, is better determined and relatively close to the BCS+HF results. 

We show in the top panel of Fig.~\ref{fig:tc_gap_ratio_src} (squares) recent gap results obtained with quantum Monte Carlo (QMC)
techniques in Ref.~\cite{Gandolfi2022}. These results are based on the Av8' interaction supplemented with the 
Urbana-IX three-body force. For Fermi momenta $k<0.4$ fm$^{-1}$, 
the agreement between the extrapolated SRC and QMC results is remarkable. QMC results peak at a somewhat lower value
of $\Delta_0 \approx 2$ MeV, a discrepancy that could be due to the difference in interactions, in many-body schemes or both.
While the intermediate density dependence is different, we note that there is an agreement in the closure Fermi momentum,
which is around $k_F \approx 1.4$ fm$^{-3}$ for both the QMC and the SRC predictions with the EM interaction. 

The central panel of Fig.~\ref{fig:tc_gap_ratio_src} shows the critical temperature obtained in the SRC approach. The maximum $T_c$
for both the EM and the EM+3NF in this approach is below $1$ MeV. This represents a decrease of $0.5$ MeV with respect to the BCS+HF data  
in the central panel of Fig.~\ref{fig:tc_gap_ratio_bcs}. This is also $0.7$ MeV lower than the maximum $T_c$ of the BCS-only
approximation. While the error bars in the plot are not negligible, with a few
points with $\epsilon_{T_c}=0.1$ MeV, they do allow for a numerical fit. The results, presented in the bottom rows of
Table~\ref{table:SRC} indicate that $T_c$ may vanish at a Fermi momentum which is around $0.05$ fm$^{-1}$ lower than
the corresponding gap closure parameter. 

Astrophysical
applications, like cooling curve simulations, rely on parametrizations of the density dependence of $T_c$, which is often 
assumed to be the same as that of the gap, $\Delta$. The results of our SRC fits indicate that the functional forms of the gap and 
the critical temperature fits may be quite different. For the EM force, for instance, the ratio of fit gap parameters is  
$k_1^\Delta/k_3^\Delta \approx 1.4$, whereas the corresponding $T_c$ fit yields
$k_1^\Delta/k_3^\Delta \approx 1.9$. A similar conclusion is drawn with EM+3NF,
thus indicating that the functional dependences are rather dissimilar. 

\begin{table}[t!]
\caption{The same as Table~\ref{table:BCS} for the SRC approximation. \label{table:SRC}}
\centering 
\begin{tabular}{cccccc} 
\hline \hline 
SRC & $g_0^{\Delta}$ & $k_0^{\Delta}$ & $k_1^{\Delta}$ & $k_2^{\Delta}$ & $k_3^{\Delta}$ \\
Units & MeV & fm$^{-1}$ & fm$^{-2}$ & fm$^{-1}$ & fm$^{-2}$ \\
 \hline
EM & 5.87 & 0.00 & 0.67 & 1.46 & 0.47 \\ 
EM+3NF & 5.04 & 0.00 & 0.57 & 1.35 & 0.39 \\ \hline
 & $g_0^{T_c}$ & $k_0^{T_c}$ & $k_1^{T_c}$ & $k_2^{T_c}$ & $k_3^{T_c}$ \\ \hline 
EM & 2.53 & 0.00 & 0.74 & 1.40 & 0.39 \\ 
EM+3NF & 2.98 & 0.00 & 0.77 & 1.30 & 0.42 \\ 
\hline 
\end{tabular} 
\end{table}

The main results of this paper are presented in the bottom panel of Fig.~\ref{fig:tc_gap_ratio_src}. We report on the ratio of the gap 
to the critical temperature within the SRC approach. The bands associated to the data incorporate the errors in both the gap and
the critical temperature, propagated assuming that they are independent to each other. 
In addition to the BCS ratio (dashed line), the figure shows the prediction $\Delta_0/T_c = 2.7 \pm 0.3$ from
Ref.~\cite{Boettcher2014}. 
This band is extracted from a functional renormalization group analysis applied to the unitary Fermi gas. This comparison is of interest
because of the un-naturally large S-wave scattering length displayed by neutrons
in the vacuum. Neutron matter can thus be interpreted as a deviation from the
standard unitary Fermi gas. 
The band obtained in Ref.~\cite{Boettcher2014} is commensurate with other predictions from different
many-body approaches. This band is also  close to a ratio of independent experimental values for $\Delta_0=0.44(3) \epsilon_F$ \cite{Schirotzek2008}
and $T_c=0.167(13) \epsilon_F$ \cite{Mark2012}, which 
yields $\Delta_0/T_c=2.63 \pm 0.27$~\cite{Randeria2014}. 

The data obtained within the SRC approach in the low Fermi momentum regime, $k_F < 1$ fm$^{-1}$, is consistent with a ratio value
of $\Delta_0/T_c=2.4 \pm 0.3$.
This is substantially larger than the BCS result, which is excluded even when considering theoretical errors in the gap 
and the critical temperature. We stress that there are substantial
differences in gap and critical temperature values between the EM and the EM+3NF, but the ratio predictions in the region 
$k_F = 0.6-1.0$ fm$^{-1}$ are extremely uniform. In particular, the ratio is insensitive to the inclusion of 3NFs. We have performed 
the same analysis for the CD-Bonn interaction, not shown here for brevity, and find also a good agreement in the ratios. 
Overall, this indicates that the ratio value due to the SRC method is robust. 

The behaviour at higher densities is more difficult to pin down. We have access to only a few data points and the 
associated errors increase substantially as $T_c$ decreases. The central values for the EM-only 
results indicate an increase in the ratio as the gap closure is reached. The results with 3NF at high densities are inconclusive, with one point
lying above and another below $2.4$.
Overall, however, there are clear indications of deviations from the BCS ratio across a wide set of densities when using the SRC
approach. 
A density-independent constant ratio could be straightforwardly implemented in cooling simulations~\cite{Yakovlev2004,Elshamouty2013}, 
before the more sophisticated density dependence shown in Fig.~\ref{fig:tc_gap_ratio_src} is explored.
The analysis of how such a renormalization of this ratio, or even how different functional forms of $\Delta_0$ and 
$T_c$ would affect astrophysical simulations, lies beyond the exploratory nature of this work.

\section{Conclusions and outlook}

In this paper, we have critically assessed the singlet superfluid gap and the superfluid critical temperature in neutron matter. 
We review whether often-used assumptions in astrophysical simulations, like the constant ratio between the 
critical temperature and the pairing gap predicted by BCS theory, hold in many-body schemes that go beyond BCS. 
To this end, we investigate two different avenues that go beyond standard BCS theory. The first one, close to the HFB approach, 
incorporates medium effects through a Hartree-Fock mean-field, $\Sigma$. The gap is informed of 3NFs through this mean-field, but
also through an averaged effective two-body contribution to the pairing interaction. We find that medium effects, incorporated in this fashion, reduce 
substantially both the
pairing gap and the critical temperature compared to the standard BCS approach. This reduction, however, does not modify the ratio 
$\Delta_0/T_c$, which remains constant across densities and very close to the BCS universal prediction. We conclude that 
the medium modifications associated to the BCS+HF approach operate effectively in the same direction in both quantities, cancelling each other.

In contrast, the SRC approach to superfluid properties considers a more sophisticated approach to medium modifications. It incorporates
an extrapolation of normal spectral functions to low temperatures and accounts for the fragmentation of single-particle strength.  
We find, in agreement with previous studies, that the pairing gap is reduced by SRCs. The critical temperature is computed here extensively  for
the first time, and we find it is also substantially reduced by SRC, compared to BCS or BCS+HF expectations. The maximum lies
below $T_c=1$~MeV across the whole density regime. Numerical fits to our data suggest that the functional form of $T_c$ as 
a function of $k_F$ may be relatively different to that of $\Delta_0$. 
In the low-density region, $k_F<1$~fm$^{-1}$, we find a universal value of the ratio $\Delta_0/T_c=2.4 \pm 0.3$.
This is substantially larger than the BCS prediction, but bodes well with theoretical simulations 
and experimental data for $\Delta_0$ and $T_c$ of the unitary Fermi gas.  

In terms of future outlook, we differentiate between three different, yet interconnected, aspects: 
developments of the SRC approach; 
quantum many-body theory extensions; and
their astrophysical impact. 
The SRC approach that we have just presented can be straightforwardly extended to other pairing scenarios and,
in fact, it has already been used to predict triplet pairing gaps at zero temperature~\cite{Ding2016,Rios2017a}. One could use this approach
to test whether the triplet gap-to-$T_c$ ratio is also modified with
respect to the BCS prediction. 
Extending the extrapolation procedure to arbitrary isospin asymmetry would also allow to explore similar effects
in the more astrophysically relevant case of $\beta$-equilibrated matter and, in particular, would allow us to provide predictions for
proton superconductors. It would be interesting to explore how NS thermal properties are modified using the SRC results presented
here. This includes, for instance, the specific heat of neutron-star matter and other important transport coefficients.

Having said that, the relatively
large errors of our numerical data are limiting the accuracy of our conclusions, and also hamper further inferences at high densities. 
Further improvements may require a new approach that actually accesses superfluid properties from the outset. 
We know, for instance, that there are missing aspects of the SRC approach that certainly modify $\Delta_0$, and can potentially
alter the $\Delta_0/T_c$ ratio. Key among these is the 
effect of long-range correlations in the pairing interaction, which was explored in Ref.~\cite{Ding2016} in a phenomenological
approach~\cite{Schulze2001,Shen2003,Cao2006} and seen to further reduce the singlet pairing gap. Moreover, the double convolution denominator in
the SRC approach does not fully account for superfluid properties and improvements of many-body theory are necessary at that level. 
Ideally, these many-body improvements should 
have associated uncertainties not only from the NN and 3NF (say, from the chiral expansion), but also from the many-body 
truncation scheme. In this sense, the ideal way forward is the recently developed Nambu covariant formalism of 
Refs.~\cite{Drissi2021a,Drissi2021b}, a theoretical formalism that fulfills all the necessary conditions to provide 
predictions in this regime. 

While quantum many-body theory developments occur, one can 
potentially explore the relevance of these results in astrophysical
simulations. The fits we provide here can be directly adopted in
cooling simulations performed with consistent EoS, to explore whether
the reduction of $T_c$ or, indeed, the change of its functional form, have any relevant observable impact. 

\begin{acknowledgements}
This work is supported by STFC, through Grants Nos 
ST/L005743/1 and ST/P005314/1; 
by grant PID2020-118758GB-I00 funded by MCIN/AEI/10.13039/501100011033;
by the ``Ram\'on y Cajal" grant RYC2018-026072 funded by 
MCIN/AEI /10.13039/501100011033 and FSE “El FSE invierte en tu futuro”; 
by
the ``Unit of Excellence Mar\'ia de Maeztu 2020-2023" award to the Institute of Cosmos Sciences,
funded by MCIN/AEI/10.13039/501100011033;
and by TRIUMF, which receives federal funding via a contribution agreement with the National Research Council of Canada.

\end{acknowledgements}

\begin{appendices}

\section{Numerical fits}\label{app:fits}

 The numerical fits have been performed with 
 Python's \verb+scipy.optimize.curve_fit+ routine~\cite{2020SciPy-NMeth},
which performs a non-linear least squares fit of data to a function.
We use the non-zero values of $\Delta_0$ and $T_c$, as well as their errors, in the fit. 
The data and fit routines are available in \verb+github+ \cite{github}. 

As a measure of the quality of the fit, we use the absolute value of 
the maximum deviation from
the data centroids to the fits, $r_\text{max}$. This is small for the 
more precise and abundant BCS+HF data, with values 
$r^{\Delta_0}_\text{max}=0.01$ MeV ($r^{T_c}_\text{max}=0.04$ MeV) and 
$r^{\Delta_0}_\text{max}=0.01$ MeV ($r^{T_c}_\text{max}=0.05$ MeV) for the EM and EM+3NF gap ($T_c$) data, respectively.
In contrast, the SRC data is more noisy and the fit has lower quality.
The maximum deviations for all quantities and potentials in this case is
$r^x_\text{max}=0.04$ MeV.

\end{appendices}

\bibliography{biblio}
\end{document}